\documentclass[11pt,twoside]{article}
\usepackage{./asp2014}

\aspSuppressVolSlug
\resetcounters

\begin{document}

\title{Three Tools to Aid Visualisation of FITS Files for Astronomy}
\author{Pragya~Mohan,$^1$ Christopher~Hawkins,$^1$, Roman~Klapaukh,$^1$ and Melanie~Johnston-Hollitt$^2$
\affil{$^1$School of Engineering and Computer Science, Victoria University of Wellington, New Zealand; \email{mohanprag,hawkinchri,roma}@ecs.vuw.ac.nz}}
\affil{$^2$School of Chemical And Physical Sciences, Victoria University of Wellington, New Zealand; \email{Melanie.Johnston-Hollitt@vuw.ac.nz}}

\paperauthor{P. Mohan}{mohanprag@gmail.com}{}{Victoria University of Wellington}{School of Engineering and Computer Science}{Wellington}{Wellington}{6140}{New Zealand}
\paperauthor{C. Hawkins}{chrishawkins.nz@gmail.com}{}{Victoria University of Wellington}{School of Engineering and Computer Science}{Wellington}{Wellington}{6140}{New Zealand}
\paperauthor{R. Klapaukh}{roman.klapaukh@ecs.vuw.ac.nz}{}{Victoria University of Wellington}{School of Engineering and Computer Science}{Wellington}{Wellington}{6140}{New Zealand}
\paperauthor{M. Johnston-Hollitt}{Melanie.Johnston-Hollitt@vuw.ac.nz}{}{Victoria University of Wellington}{School of Chemical and Physical Sciences}{Wellington}{Wellington}{6140}{New Zealand}

\begin{abstract}
Increasingly there is a need to develop astronomical visualisation and manipulations tools which allow viewers to interact with displayed data directly, in real time and across a range of platforms. In addition, increases in dynamic range available for astronomical images with next generation telescopes have led to a desire to develop enhanced visualisations capable of presenting information across a wide range of intensities. This paper describes three new tools for astronomical visualisation and image manipulation that are the result of a collaboration between software engineers and radio astronomers. The first tool, FITS3D, is a fast, interactive 3D data cube viewer designed to allow real-time interactive comparisons of multiple spectral line data cubes simultaneously. It features region specific selection manipulation including smoothing. The second tool, FITS2D, aids the visualisation and manipulation of 2D fits images. The tool supports the interactive creation of free-form masks which allow the user to extract any (potentially non-contiguous) subset of a fits image. It also supports annotations which can be placed without affecting the underlying data. The final tool is an R package for applying high dynamic range compression techniques to 2D fits images. This allows the full range of pixel brightness to be imaged in a single image, simultaneously showing the detail in bright sources while preserving the distinction of faint sources. Here we will present these three tools and demonstrate their capability using images from a range of astronomical images. 

\end{abstract}

\section{Introduction}
As we move to larger and more data rich astronomical images there is a pressing need to develop fast, high quality interactive visualisation tools which can deal with the increased images sizes expected from next generation astronomical instruments. In particular, there is a lack of capability in interactive rendering of 3D FITS cubes, extraction tools for 2D FITS files and compression for high dynamic range astronomical images. Here we briefly present three such tools for the astronomical community. 

\section{3D rendering: FITS3D}

There are a number of tools which allow for viewing of 3D images (e.g. 
DS9;~\citet{joye2003} and Karma;~\citet{gooch1996}). 
Historically, these tools are often slow or cannot be used easily on a commodity
laptop\footnote{For example, DS9 took 22 seconds to finish rendering a
single frame of a 270Mb FITS file.}. Recently \cite{hassan2011} described
a system which can open a 26Gb FITS files viewed at 3 frames per second 
using a set of networked computers and AstroVis~\citep{perkins2014} is able 
to render a 1.8Gb FITS file at 40 frames per second. However, even these newer 
systems are slow to load files\footnote{AstroVis takes more than 5 minutes to 
load a comparatively small 272Mb FITS file.} and often run a lower speeds 
(down to 1 frame per second) when the user zooms into an image.
For the next generation of large telescopes such as the Square Kilometre Array, 
we need to be able to open images of many 10s of GBs, for at least preliminary 
viewing on a commodity laptop.

FITS3D\footnote{\url{https://github.com/chrishawkinsnz/Fits3D}} has been
created to allow for fast and interactive exploration of
multiple 3D and 4D FITS files simultaneously. FITS3D reads in a FITS file with 
user controlled quality
settings. At the lowest quality we are able to read in a 26Gb FITS file
in 0.78 seconds, and then display it at approximately 280 frames per 
second\footnote{Using a 2014 MacBook Pro.}. In addition to standard imaging
options such as real time histogram-based filtering, smoothing and rotations, FITS3D 
also allows rapid changes to quality settings for either the entire space, or only 
a user specified area (Figure~\ref{fig:fits3d}). All of these mechanisms
are optimised to both allow the clearest view of the data and ensure
the best possible performance. As not all programs are able to load 
cubes of the same size as FITS3D, it is also able to export user selected
cubes, which can then be used by other programs. 

It is also possible to load in multiple cubes simultaneously using 
FITS3D. If these have the same coordinate systems, then the cubes will
be overlaid using shared coordinates, so that two cubes can be looked at
simultaneously and their intersections can be clearly (and correctly) 
seen. FITS3D also supports the exploration of 4D hypercubes, where the 
fourth dimension can be stepped through using a slider. 

\articlefiguretwo{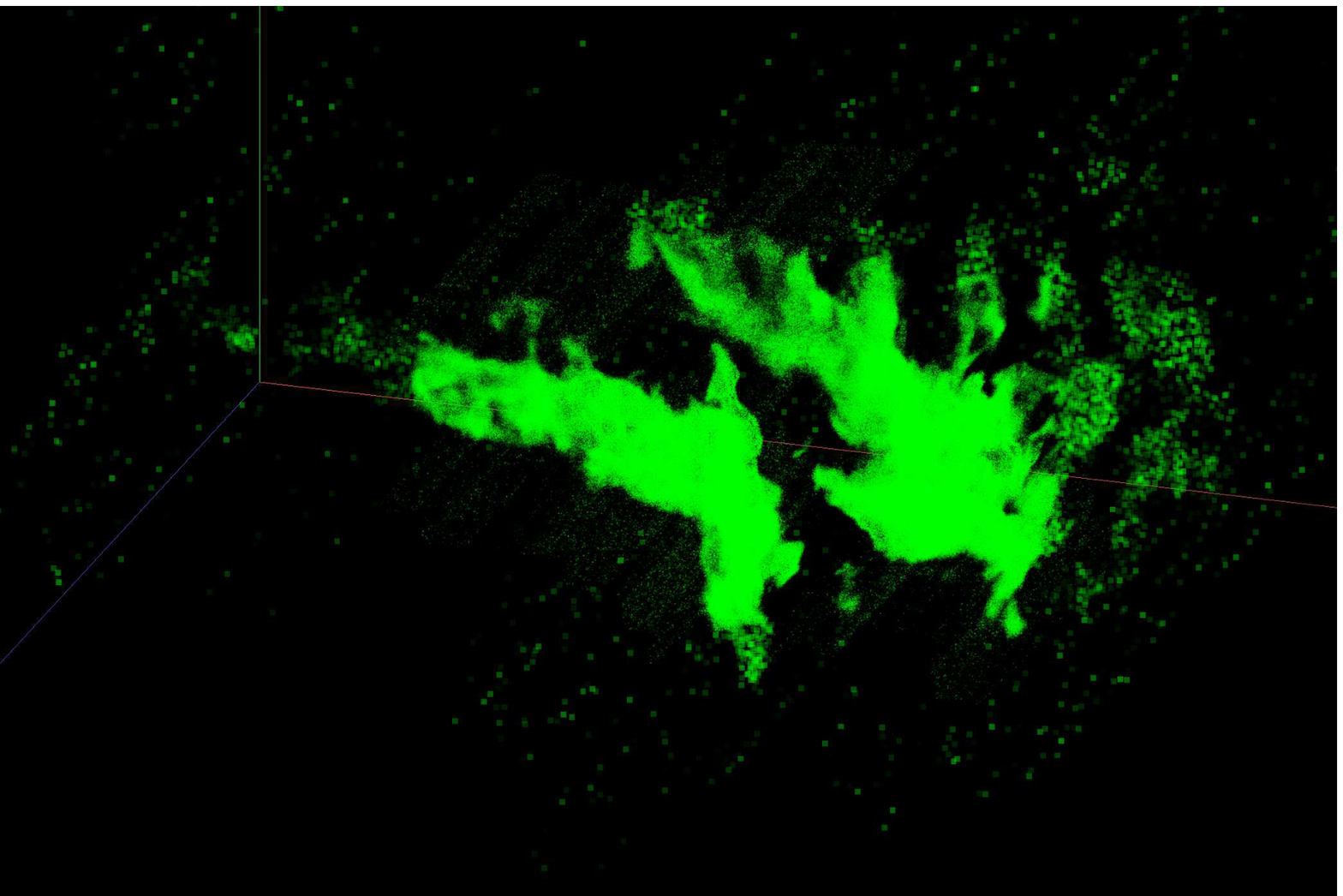}{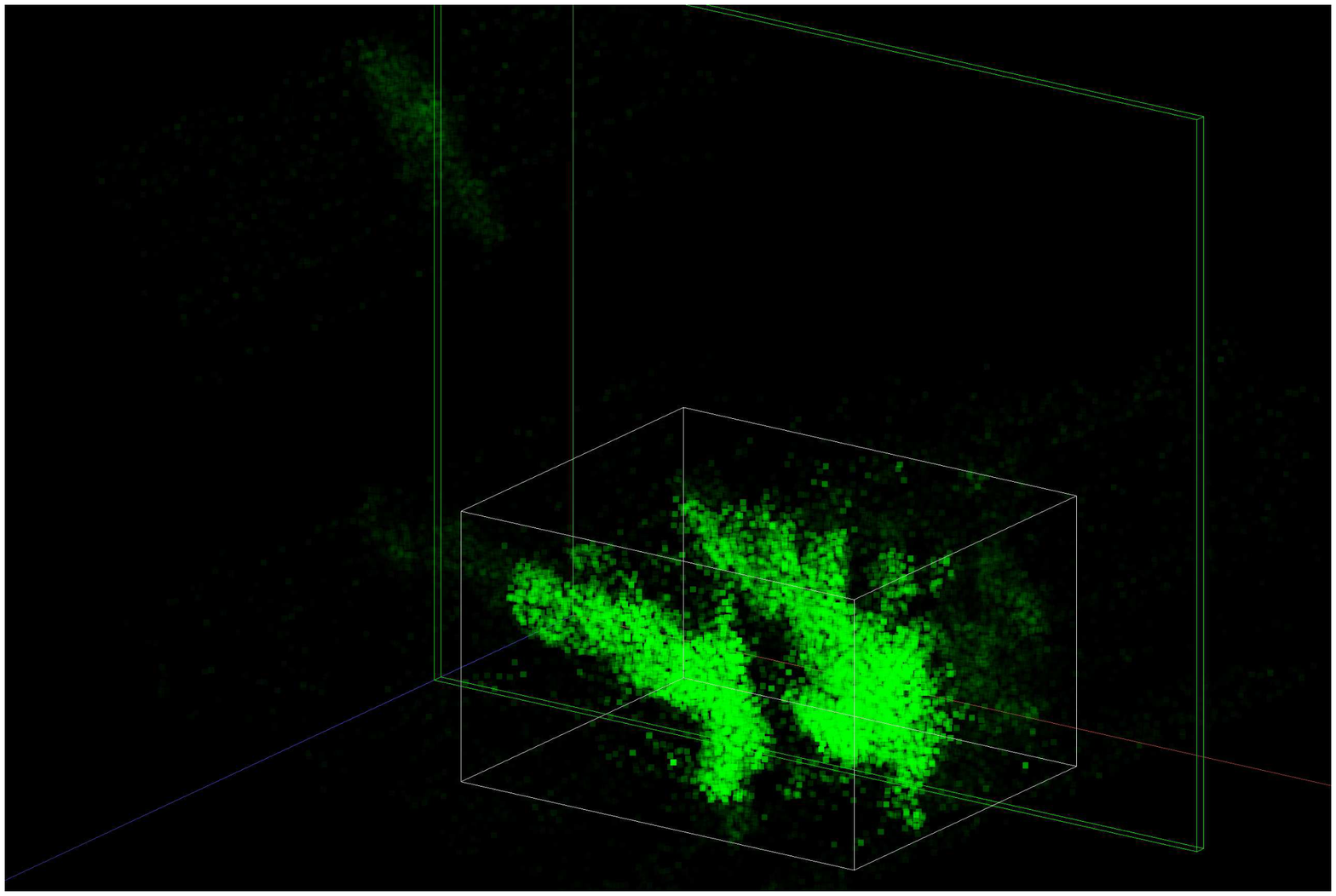}{fig:fits3d}{3D cubes rendered using FITS 3D \emph{Left:} An image with different quality for different regions.  \emph{Right:} User selection box.}

\subsection{Region Extraction \& Annotation: FITS2D}

The best current option for extracting region from a FITS file is the 
MIRIAD toolkit~\citep{sault1995}, which allows the user to extract a polygon
with hand written coordinates. Alternative tools such as the Fits Cutout 
web service \citep{haridas2005} are even more limited, only allowing the 
extraction of rectangular regions. However, astronomical objects are
frequently complex shapes, often with occlusions or non-continuous parts 
which are hard to select cleanly this way. Additionally, region annotation 
with text and simple shapes like ellipses and rectangles is
a feature that exists in a number of software packages such as 
kvis~\citep{gooch1996} and DS9, but the ability to create annotations of an arbitrary shape 
is lacking. 

\articlefiguretwo{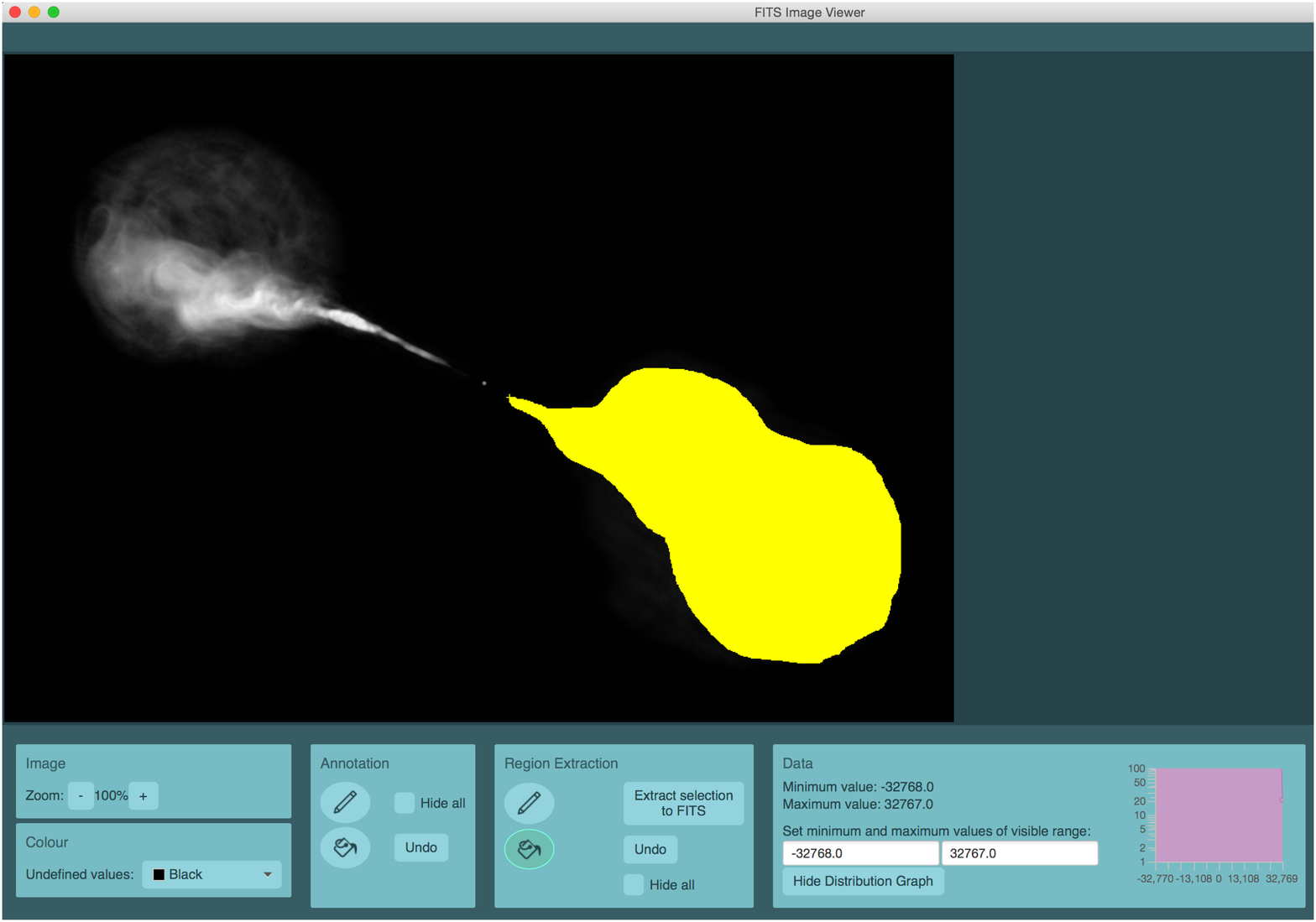}{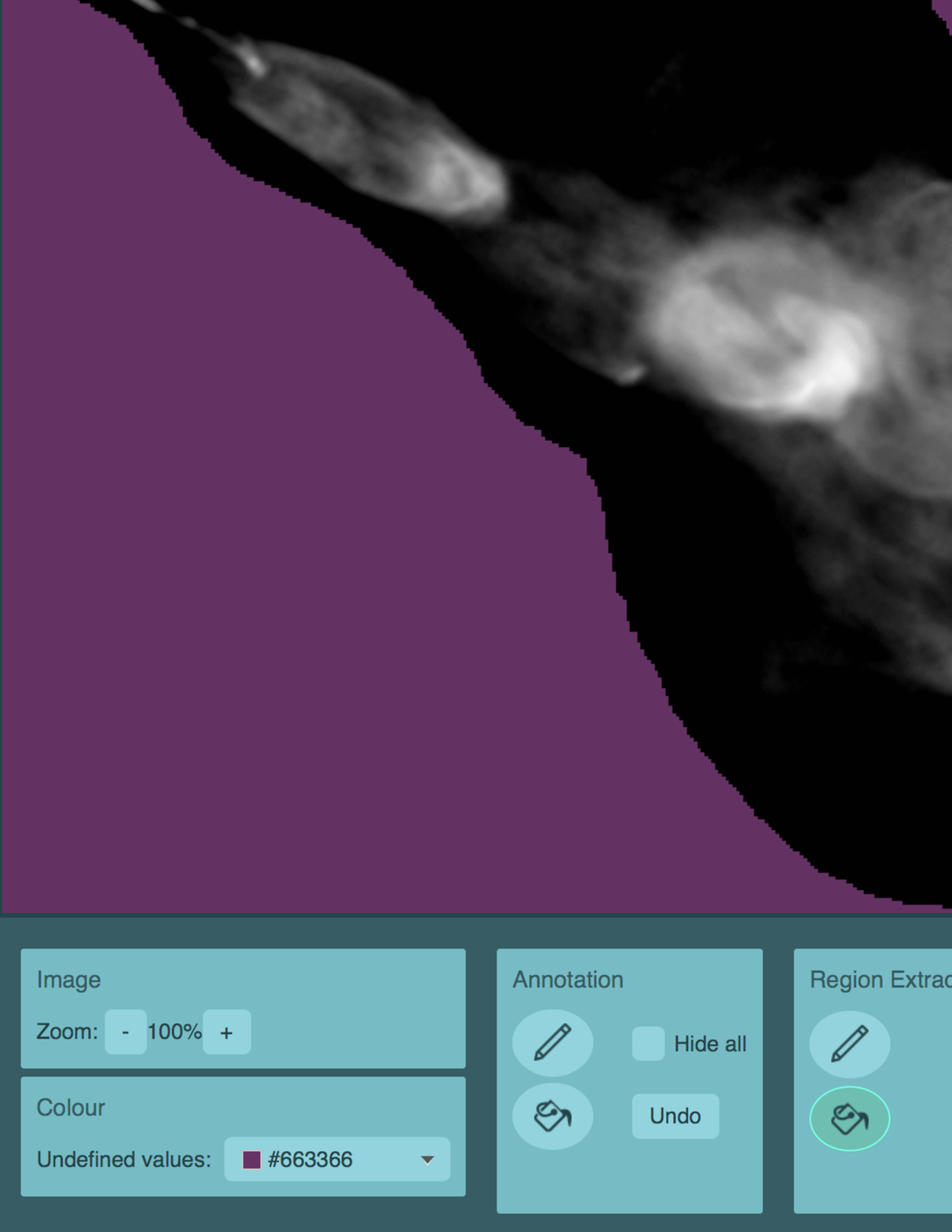}{fig:fits2d}{An example of region extraction using FITS2D \emph{Left:} Object of interested being painted by the user  \emph{Right:} Object of interest now loaded as a separate file (note that pink represents blanked values).}

FITS2D\footnote{\url{https://github.com/PragyaM/FITS2D}} is a tool to support
the existing collection of 2D FITS image viewing and manipulation tools. 
Its primary features are free form region extraction and annotation. Both extraction and 
annotation share a single mechanism based on standard image
painting applications. A user is able to draw a freehand path using the paintbrush 
tool and may fill any closed shape using the paint bucket tool. Pixels selected either by the
paintbrush or bucket tool may then be extracted as a new FITS file (see Figure~\ref{fig:fits2d}). 
Since the FITS format expects are rectangular image, the generated image will be a 
rectangle larger than the selected area, but any non selected pixels will simply 
be exported as missing values. Annotations are the same, except that annotations
can also be exported to a separate file (including their coordinates), and then reimported
into any other files. The coordinates stored are used to ensure that the annotations
appear in the same spatial location in all files.

\subsection{High Dynamic Range Image Compression}

High dynamic range (HDR) compression is the general problem of trying
to show a large range of values when you only have a small number
of output values. Gradient domain high dynamic range 
compression~\citep{fattal2002} attempts to address this problem by 
compressing the derivative of the image, then reintegrating (numerically)
in order to get back a compressed image. This technique has been
applied extensively to photography, but is also applicable to astronomical
FITS files. We explored the use of gradient domain high dynamic range
compression on FITS images with the help of an R
library\footnote{\url{https://github.com/klapaukh/astror}} and a C 
implementation\footnote{\url{http://ttic.uchicago.edu/~cotter/projects/hdr_tools/}}  
of the algorithm.


Figure~\ref{fig:hdr} shows there is some difference between the simple
log scale and using the HDR compression. However, we find that the HDR compression is
significantly slower (in large part due to the numerical integration required). We do
note that there are other techniques for HDR compression that can be applied to
astronomical images, and we hope to see more work done on this in future.


\articlefiguretwo{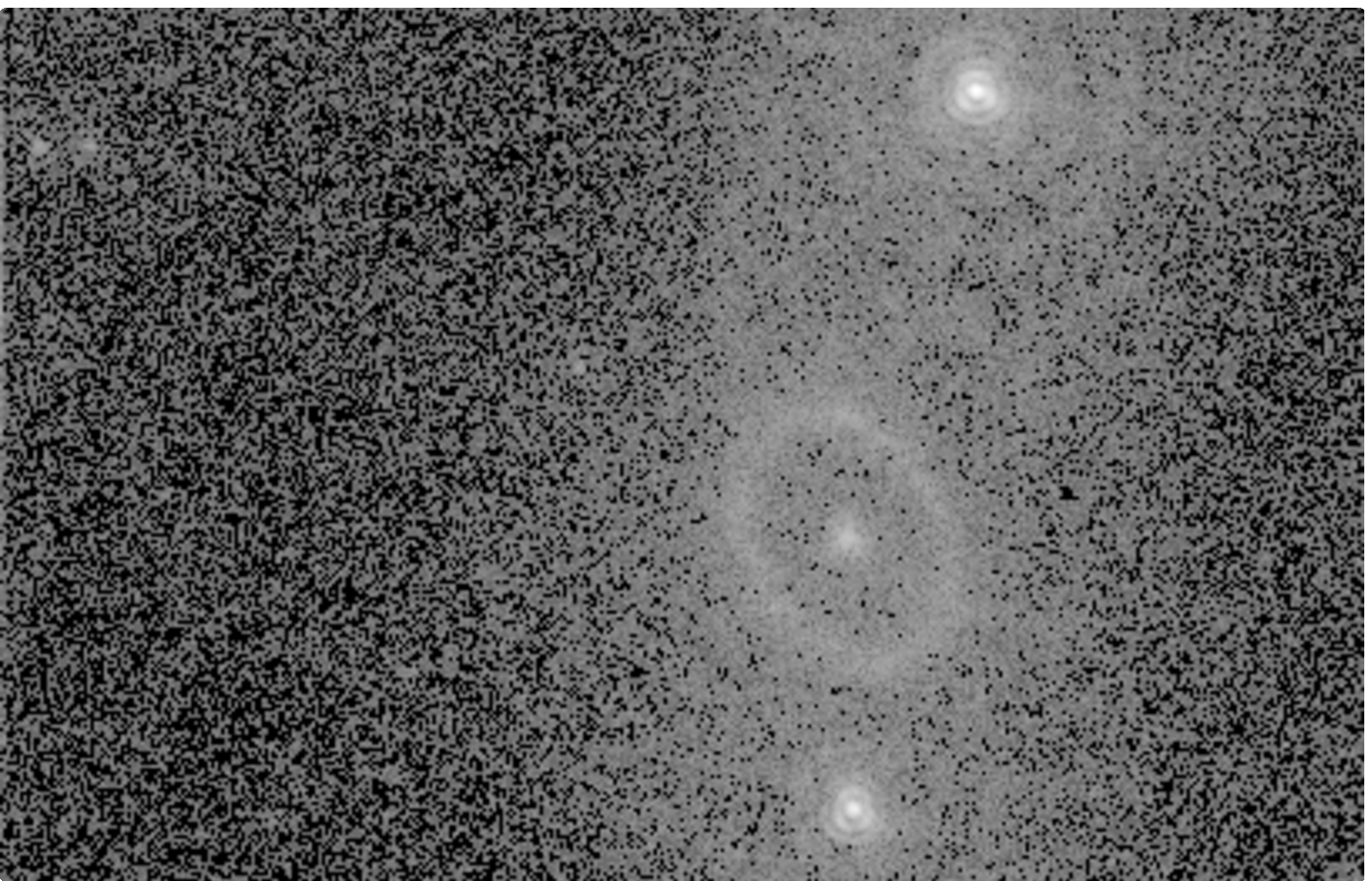}{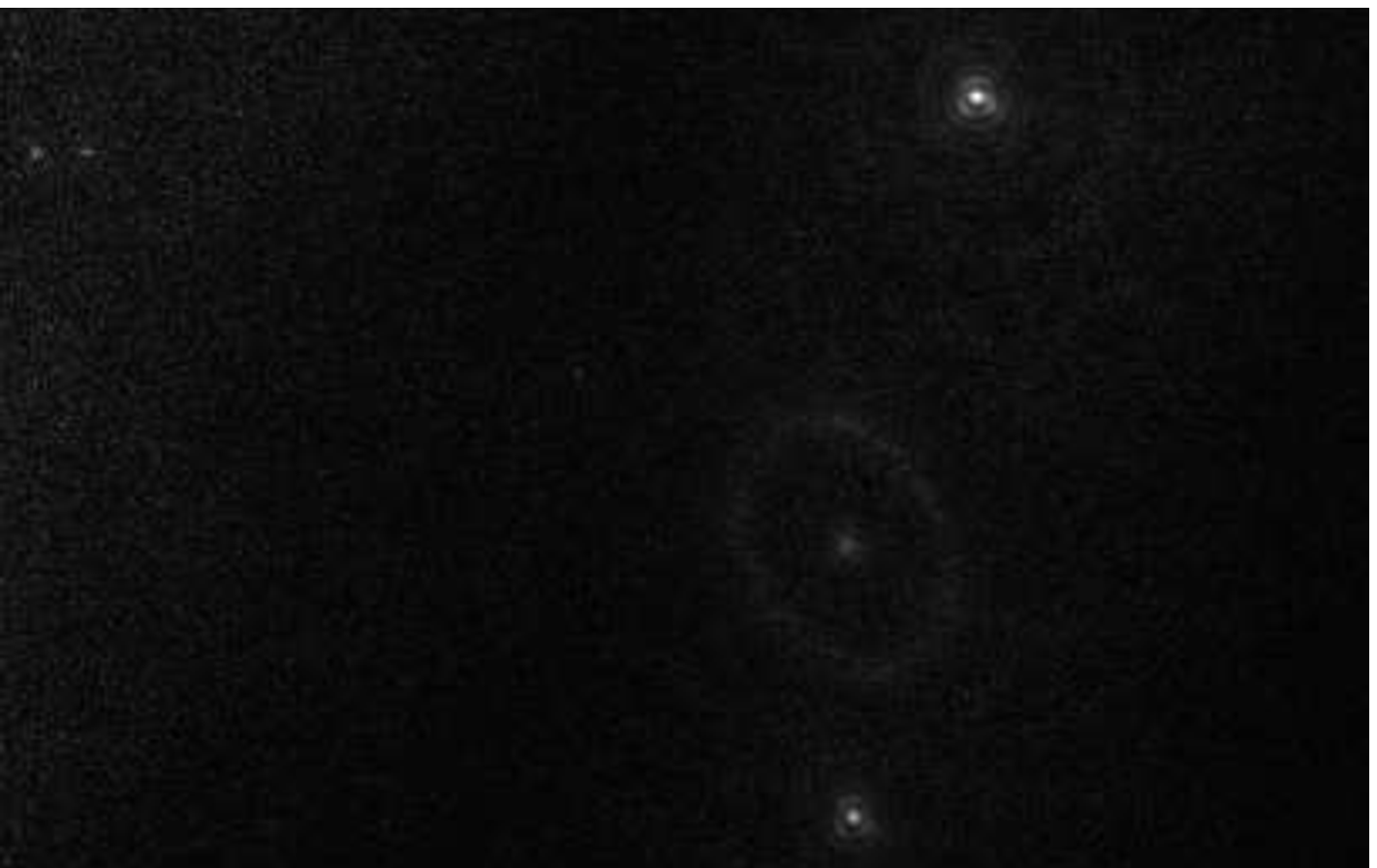}{fig:hdr}{FITS test image from \url{http://www.cv.nrao.edu/fits/}. \emph{Left:} Log scaled. \emph{Right:} Using the HDR compression.}

\section{Conclusion}
We have developed three new tools for exploration of astronomical data. Two of these tools, FITS3D and FITS2D, will be described in detail in future publications and all three packages are available for beta testing in the repositories given here.

\acknowledgements MJ-H is supported in this work by a Marsden Grant. We thank L. Pratley \& L. Hindson for discussions on limitations of existing astronomical tools. 

\bibliography{O10.2}  
\end{document}